\documentclass[a4paper,11pt]{article}
\pdfoutput=1
\usepackage{jinstpub}

\usepackage{graphicx}

\newcommand{\cm}{\ {\rm cm}}

\newcommand{\GeV}{\ {\rm GeV}}
\newcommand{\MeV}{\ {\rm MeV}}
\newcommand{\nC}{\ {\rm nC}}
\newcommand{\nm}{\ {\rm nm}}
\newcommand{\Hz}{\ {\rm Hz}}

\title{Detection of Dark Photon Decaying into $e^+e^-$ using Cherenkov Radiation}

\author[a]{Suyong Choi \note{Corresponding Author}}
\author[a]{Yeonjun Kim}
\author[a]{Youn Roh}

\affiliation[a]{
Department of Physics, Korea University \\
Seoul 02841, Republic of Korea
}
\emailAdd{suyong@korea.ac.kr}
\emailAdd{yunjun3@naver.com}
\emailAdd{youn@cern.ch}

\date{\today}

\abstract{
In dark photon search experiments with electron beam-dumps, it is difficult to access the smaller dark photon life-time region of phase space due to enormous backgrounds from low-energy particles emerging from the target. In order to reduce the background, a thick beam-dump target is usually necessary.
We propose to detect the Cherenkov radiation in gas due to ultra-relativistic electron and positron from dark photon decay.
The secondary particles emerging from the beam dump have very little chance to produce such Cherenkov radiation in gas. Making use of the  direction of the Cherenkov radiation, low background dark photon search with thinner target is possible. This would allow one to access challenging regions of the dark photon parameter space with low power electron beams and low-cost experimental setup.
}

\keywords{Dark photon, Beam dump experiment, Cherenkov detector}
\begin{document}
\maketitle


\section{\label{sec:level1}Introduction}

Understanding the nature of the dark matter content of the universe is
one of the most important problems in particle particle physics. 
Models with massive dark photon, an extra $U(1)$ gauge boson
with weak coupling to SM charged particles, are a bottom-up approach to its resolution with
interesting possibilities \cite{Pospelov}. 

Many experiments had searched for an exotic particle in the MeV to GeV mass range 
that decays into an electron-positron pair, without any significant signal so far. 
Null results from these experiments have been reinterpreted and large parameter space have already been excluded, 
especially for the long-lived dark photons  (Fig. \ref{fig:reach})  \cite{review}. 
Unexplored region of parameter is quite challenging to probe experimentally and
new techniques should be found. We propose to detect Cherenkov radiation
as a clean signature of dark photon decay.

In this letter, we consider the case where the dark photon decays into a pair of electron and positron.
We assume that the candidate for stable dark 
matter (DM) is massive enough that the dark photon does not decay into a
pair of DM particles. In this case, the property of dark photon is fully
determined by its mass ($m_{A'}$) and a coupling parameter ($\epsilon$) \cite{Bjorken}. 
The dark photon couples to standard model charged particles
with coupling strength $\epsilon e$. For $m_A' < 2m_\mu$, $A'$ decays exclusively into $e^+e^-$.

Production of $A'$ in electron beam dump is similar to
the Bremsstrahlung process \cite{Bjorken}. 
If electron beam energy ($E_{beam}$) is much greater than $m_{A'}$, the energy of the $A'$ produced 
is close to that of the incoming electron beam. Also, $A'$ is highly directional,
forming a very small angle $\theta \ll m_{A'}/E_{A'}$ with respect to the incident electron beam direction.
If the dark photon decays after emerging from the beam dump, the 
visible decay particles can be detected. For a long-lived dark photon, a way of detecting 
its decay would be to sum the electron and positron energies. This should be close to $E_{beam}$ \cite{Bjorken}. 

\section{Deficiency of Calorimetric Method of Dark Photon Detection}

A method of detecting the $A'\rightarrow e^+e^-$ with long decay length involve 
calorimeters for energy measurement of electron and positron. This method is
adequate for detecting very long decay lengths of $A'$, since with thick enough target and shielding, 
almost all secondary particles can be stopped.

For promptly decaying dark photons, precision spectrometer combined
with thin targets are used. Due to limited acceptance of spectrometers, they are not
as well suited to detect long-lived $A'$ decays.
In order to access the unexplored region of intermediate lifetime region of parameter space,
the detector needs to be closer to the target and the target needs to be thinner as well.
A good example of this is the Heavy Photon Search (HPS) experiment at JLAB, which uses vertex detectors 
in addition to calorimeters. The HPS experiment is sensitive to both prompt and events with displaced vertex \cite{HPS}. 
However, due to low rate of interaction in thin targets, accelerator with large beam current
is required.

\begin{figure}
\centering\includegraphics[width=0.8\linewidth]{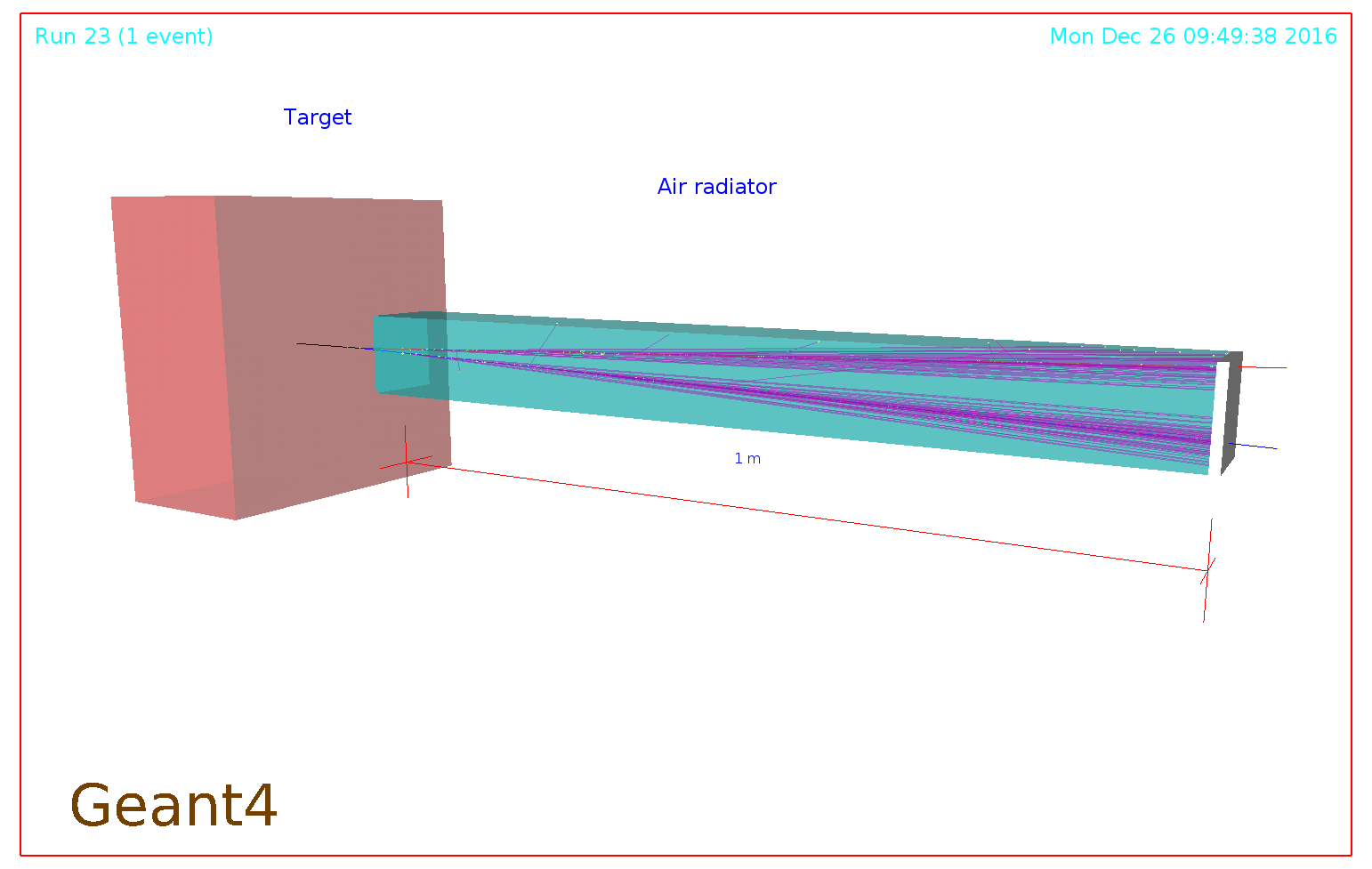}
\caption{\label{fig:aircherenkovevent} GEANT4 simulation of a dark photon event $A'\rightarrow e^+e^-$ with $E_{A'}=3\  {\rm GeV}$, $m_{A'}=0.1\  {\rm GeV}$, $\epsilon=2\times 10^{-5}$. The shaded narrow column is the air radiator. Shields surrounding the radiator have been omitted to permit better view. The two clusters of lines are due to Cherenkov photons produced by 
relativistic electron and positron.}
\end{figure}

In accelerators with lower current, we must use thicker targets or
run for longer periods to be competitive. 
For targets of intermediate thicknesses, copious low-energy secondary particles
would be produced.  To be able to track and measure the electron and positron signal
would require fine tracking and calorimetry, which increases the cost and complexity. 
Careful optimizations are need to make it feasible. Obviously, use of a very thick target
would eliminate many secondaries, but this region is already excluded by other experiments.

In the next section, we outline a method that allows us to use a target of
intermediate thickness. We will exploit the fact that
the secondary particles emerging from the target have extremely small chance to
produce Cherenkov radiation in gases, while electron and positron from 
$A'$ decay produce Cherenkov radiation in gases with high probability.

\section{Detection of $A'\rightarrow e^+e^-$ from Cherenkov Radiation}

Cherenkov photons are produced in air When a relativistic charged particle of $\gamma>40$ travels 
at standard temperature and pressure (STP) whose index of refraction is 1.0029 for air.
These photons are highly directional, forming a small angle with respect to the charged 
particle $\theta=\cos^{-1} \frac{1}{n\beta}=0.024$.
Such a relativistic charged particle traveling through 1 meter of air will produce about 90
photons in the wavelength range of $200$--$700\nm$. This number gets reduced to less than half
for wave lengths in the $200$--$300\nm$ range.
Since the speed of light in air at STP is very close to that in vacuum, the
Cherenkov photons that are continually produced as the particle travels, will occupy
the similar longitudinal position. Therefore, the arrival
time of the photons at the far end would be almost coincident with little spread.
We can exploit the directional and coincident nature of the Cherenkov photons.

There are two sources of backgrounds that need to be considered. 
The first one is scintillation from nitrogen gas in the air. When low energy particles
that emerge from the target lose their energies in the air volume, nitrogen molecules emit
scintillation light during de-excitation. The nitrogen scintillation is in the range of $300$--$400\nm$ \cite{nitrogenscint1,nitrogenscint2}.
The lifetime of nitrogen gas scintillation is on the order of a nanosecond at STP, while the Cherenkov radiation is prompt. However, separating the scintillation photons from Cherenkov photons using
timing will not be possible. One distinguishing feature is that the scintillation photons 
are not directional, while Cherenkov photons are.
By exploiting this fact, Nitrogen scintillation background can be suppressed
with suitable optical reflection system that directs Cherenkov photons with
high efficiency towards photodetector.
We can further supress this background by lowering nitrogen content, for example, 
by introducing carbon dioxide gas. Another way to 
reduce sensitivity is to use UV-sensitive photomultiplier tubes. 

Another source of backgrounds are energetic electrons that can produce Cherenkov radiation.
If the target is not thick enough,  electromagnetic shower
is not fully contained and electrons whose energy is above the  
$40\MeV$ threshold, may escape the target. This limits the thinness of the target.
Another way Cherenkov radiation can be produced is
if high-momentum proton knocks electrons free inside the radiator volume or the surrounding shielding.
Although extremely rare, protons can be knocked free from the target material by incoming electron beam.
If a pulse of electron beam contains large number of electrons, it may release large number of energetic
protons. If the delta rays created by proton has enough energy, it can produce Cherenkov photons. These
Cherenkov photons, however, may form larger angles with incident electron beam direction.,
 due to the wide spread of delta electron directions.

\begin{figure}
\begin{center}
\includegraphics[width=0.6\linewidth]{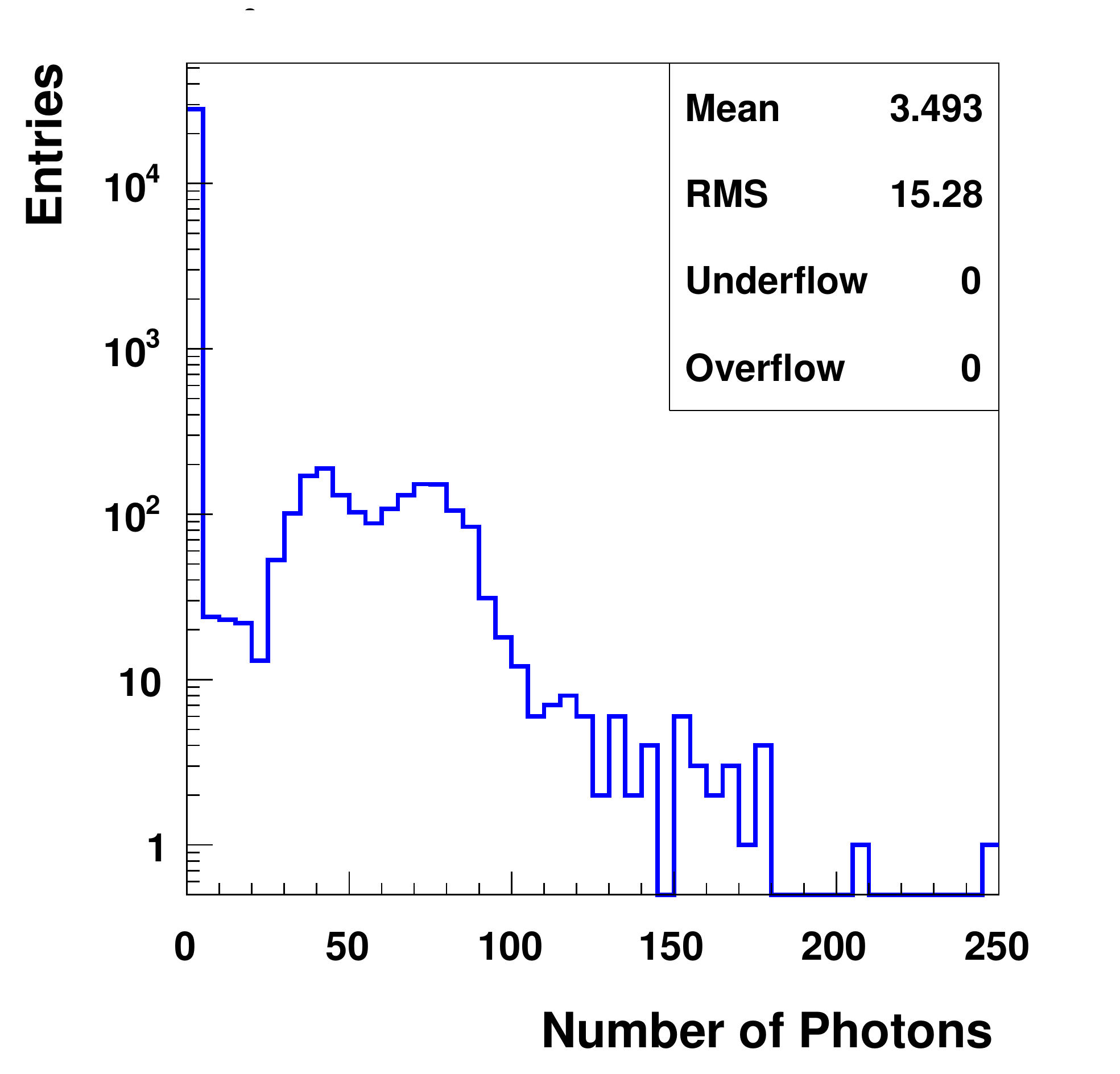}
\end{center}
\caption{\label{fig:optphotons} Distribution of number of optical photons from 10000 dark photon signal events reaching the detector for $m_{A'}=0.1\ \mathrm{GeV}$ and $\epsilon=2\times 10^{-5}$.}
\end{figure}

In the next section, we examine the potential reach of this method with a conceptual experiment using a low current 3 GeV electron linear accelerator (LINAC) in Korea. With this example, we demonstrate the potential benefits of this method.

\section{An example case: Dark Photon Search Experiment Concept at Pohang Accelerator Lab}
The Pohang Light Source II (PLSII) at Pohang Accelerator Laboratory (PAL) in Korea is a 
synchrotron radiation facility \cite{PALPLS}.
A 3 GeV electron LINAC supplies the electrons to the circular electron storage ring 
in a top-up mode in order to keep the current constant in the storage ring. 
The LINAC produces electron beam bunches of $0.5\nC$ each with $10\Hz$ repetition rate.

\begin{figure}
\begin{center}
\includegraphics[width=0.6\linewidth]{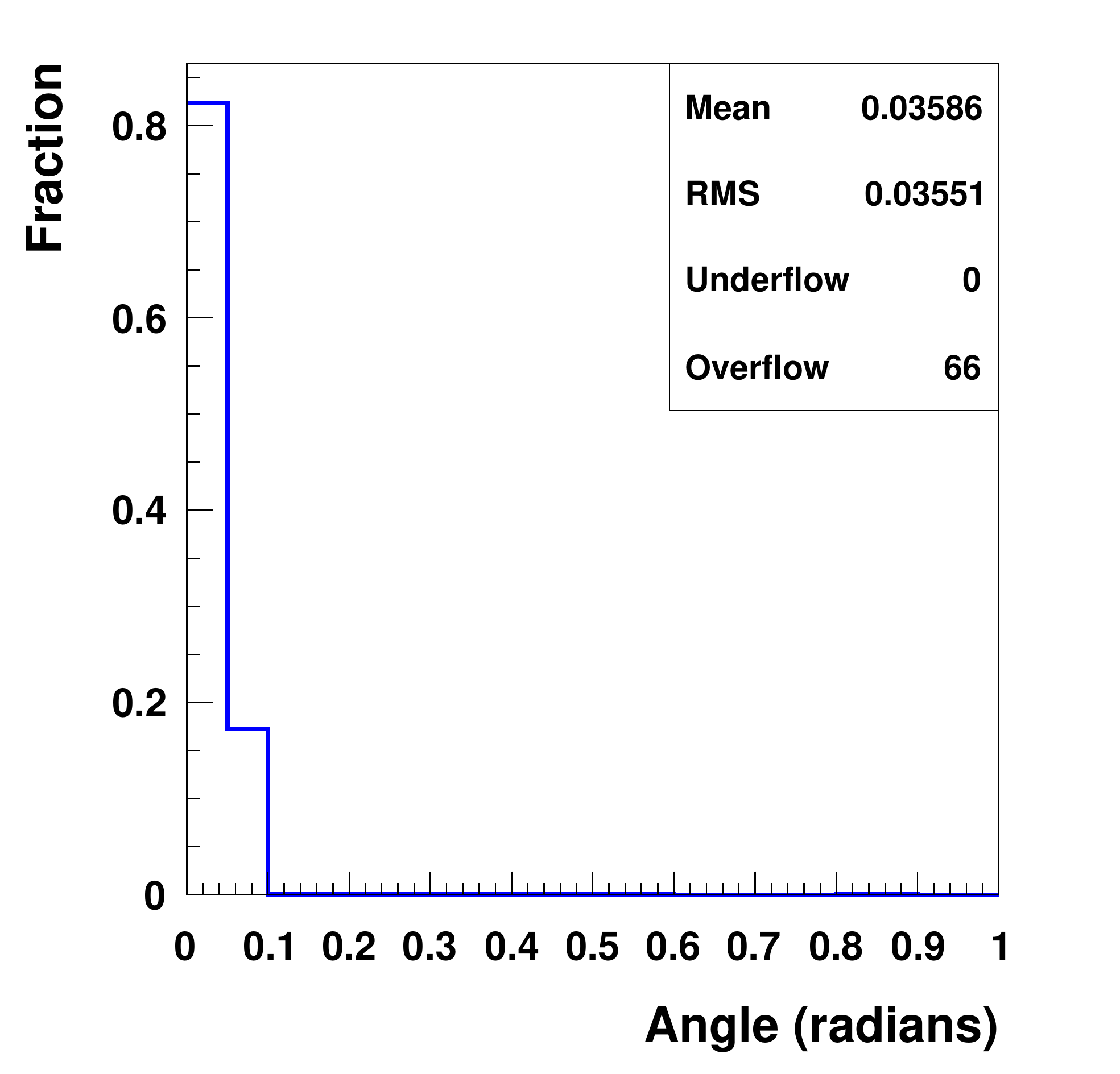}
\end{center}
\caption{\label{fig:aircherenkov-darkphoton} Distribution of incident angle due to photons with wavelengths in $200-300\nm$ range from dark photon signals for $\epsilon=2\times 10^{-5}$. Histograms are normalized to unity.}
\end{figure}

The PLSII has a beam dump where a dark photon search experiment can be performed
when the storage ring is not in use. Compared to other electron beam LINAC, the PLSII LINAC is not the most powerful machine. However, one advantage of the PLSII is its low repetition rate.
This means that the slow secondary particles emerging from the beam dump do not affect the measurement if the timing information is used. The Cherenkov method can turn this moderate
facility into a competitive site for dark photon search.

In order to test the feasibility of the Cherenkov radiation method, we use GEANT4 detector simulation library, which contains
the accurate physics of particle and nuclei interactions in matter \cite{geant4}. We used the known electromagnetic
and hadronic interactions through the ``QGSP\_BERT'' physics list. To investigate the optical photon production
and propagation in air, we used the Cherenkov radiation physics and implemented the nitrogen gas
scintillation, based on the measurements from real data. The dark photon decay physics was implemented
to study the acceptances \cite{nitrogenscint1,nitrogenscint2}.

We consider a $20\ {\rm cm}$ thick beam dump target made of tungsten, which has a high density 
and short radiation length of $X_0=0.32\ {\rm cm}$. 
However, this is not enough to stop all the secondaries produced in the target.
For each incident $0.5\ {\rm nC}$ electron beam, the sum of energy of the particles emerging from the target amounts to a few TeV's.
Although probability for secondaries to emerge from the target is very low, with enough incident electrons, the total amount could be sizable.

In contrast, dark photon signal would deposit an energy of $3\ {\rm GeV}$, which is
quite small compared to the level of backgrounds. Using calorimetry alone would make it 
difficult to separate the signal from the backgrounds. In this case, a thick target is needed and 
access to the smaller lifetime parameter space becomes difficult.
Fine grained calorimetry together with fine-grained tracking might mitigate the effects, but
a careful study is necessary to understand the issues of high occupancy.

\begin{figure}
\centering\includegraphics[width=0.6\linewidth]{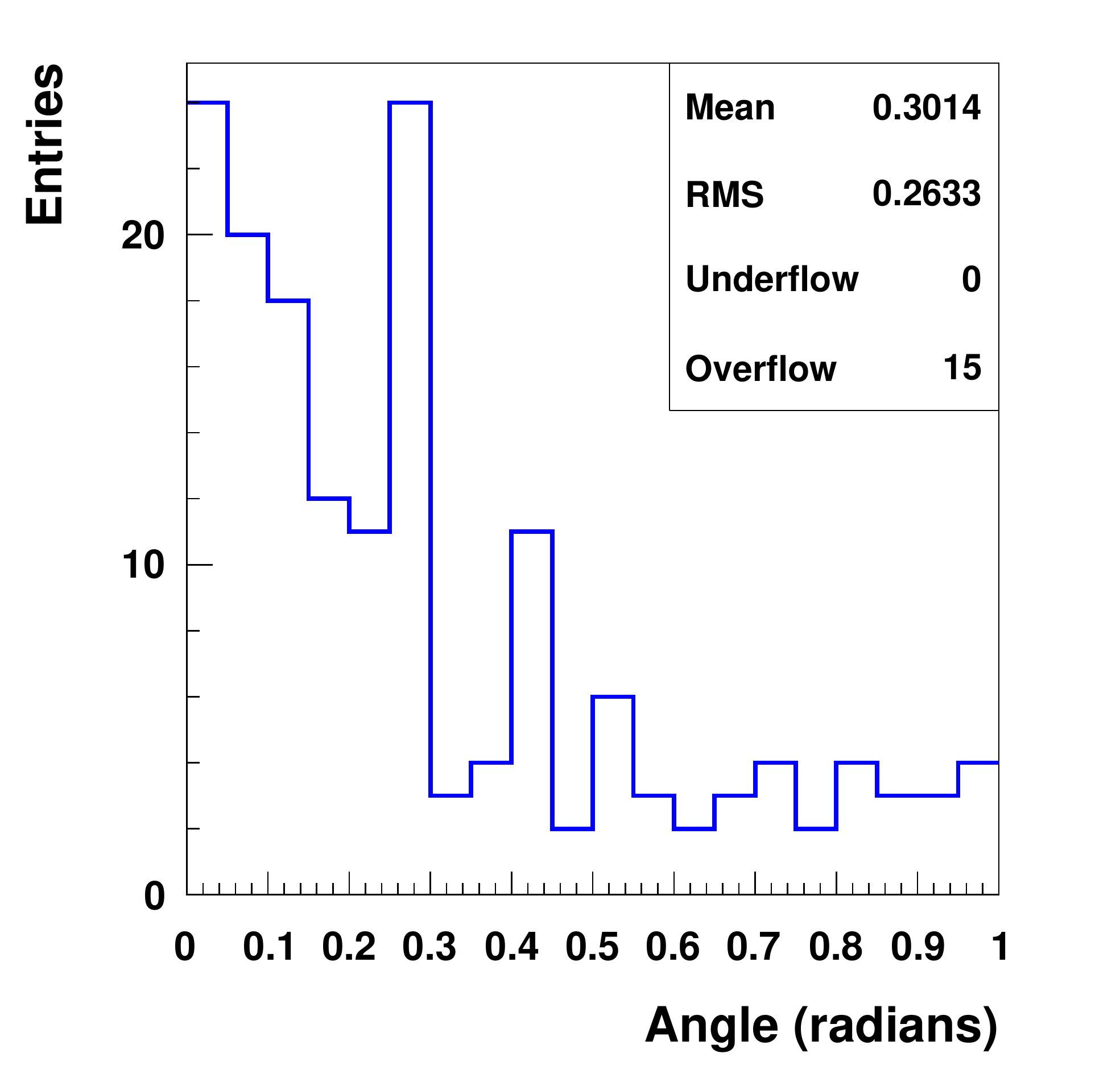}
\caption{\label{fig:timeangle} Angular distribution of optical photons reaching the detector for $5\times 10^8$ 3 GeV electrons on the tungsten target.}
\end{figure}

Instead of a calorimeter, we place air Cherenkov radiator, whose dimension
is $10\times 10\cm^2$ transversely and $100\cm$ longitudinally. 
The volume of air is surrounded by $1\cm$ thick tubular tungsten shielding. 
At the far end of this radiator, we place a
pseudo-detector to measure the  angle of incidence of optical photons. For a realistic detector,
one should place a mirror at the end of the radiator and reflect the optical photons onto a strategically 
placed photo-detectors.
A dark photon that would emerge from the target subsequently
decays into a $e^+e^-$ pair. The electron and positron, each has $1.5\GeV$ in energy, and whose 
gamma factor $\gamma\approx 3000 \gg 40$
is large enough to produce Cherenkov photons while traveling through air (Fig. \ref{fig:aircherenkovevent}). 

The distribution of the number of optical photons reaching the end of the air column is shown in Fig. \ref{fig:optphotons} for $\epsilon=2\times 10^{-5}$ and $m_{A'}=0.1\GeV$ from 30000 dark photon events. We restrict the wavelengths of photons from $200\nm$ to $300\nm$, since we want to block scintillation photons.The dark photon parameter chosen corresponds to a dark photon mean decay length of $6\cm$.
Due to the short decay lengths, about 94\% of the events do not produce significant signal. 
For smaller values of $\epsilon$, the efficiency rises as more dark photons decay
behind the beam dump. For much smaller values of  $\epsilon$, where the decay length
becomes longer than the Cherenkov radiator, efficiency starts to decrease.

Two broad peaks above 20 are visible in Fig. \ref{fig:optphotons}. 
The first peak is due to events where only one of electron or positron 
travels through the volume and produces Cherenkov photons that reach the detector. The second peak
corresponds to the case when Cherenkov photons from both the electron and positron
reach the detector. A larger radiator in the transverse direction would allow one
to capture all the photons. 

Figure \ref{fig:aircherenkov-darkphoton} shows the directional property of the Cherenkov photons that reach the pseudo-detector.
Nitrogen scintillation photons, on the other hand, do not have preferred directions. 
Therefore, this property of Cherenkov photons can be exploited by using optical systems to select only the forward-going photons,
in a realistic experiment.

In order to estimate the backgrounds we inject $5\times 10^8$ electrons on the tungsten target in GEANT4 simulation.
This is about one sixth of the number of electrons in 0.5 nC electron beam pulse. 
Despite the large number of electrons hitting the target, only 163 photons whose wavelengths are in $200$---$700\nm$, reach the detector.
The distribution of angle of incidences of these photons are shown in Fig. \ref{fig:timeangle}. 

Most of these photons are due to nitrogen gas scintillation, while 14 photons under the peak at 0.3 radians 
are due to a single instance of an electron producing
Cherenkov radiation. If we restrict to wavelengths in $200$---$300\nm$, only the event in the peak would remain
and the number of photons would be reduced to 7. 
To be insensitive to the nitrogen gas scintillation photons,  one should use either short pass optical filter and/or a photo-cathode of a photo-detector that is sensitive in the ultraviolet range and insensitive  to wavelengths above $300\nm$. 
Another way to achieve this is to use a different gas, such as carbon dioxide, which doesn't produce photons in the optical range.

\begin{figure}
\centering\includegraphics[width=0.6\linewidth]{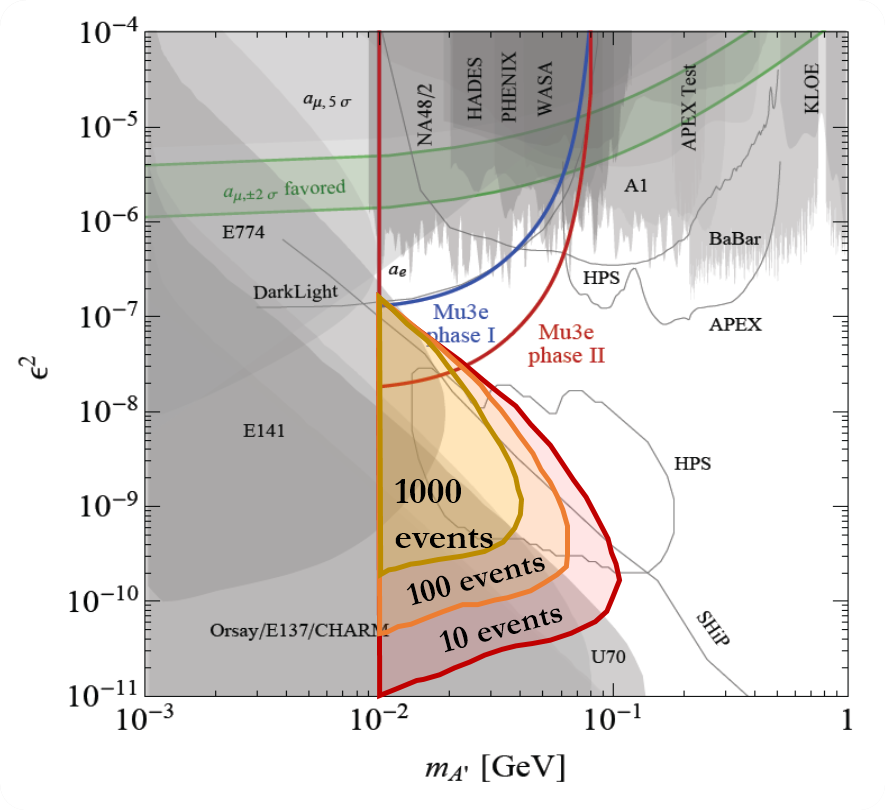}
\caption{\label{fig:reach}Reach of dark photon search experiment using 3 GeV electron beam dump at PAL PLSII with seven days-equivalent of data-taking.}
\end{figure}

If we assume that the Cherenkov photons produced by delta rays follow the angular distribution of the scintillation photons,
 we can expect 10 Cherenkov photons on average to fall within $0.1$ radians. Probability for 10 photons to fluctuate to 30
 or more photons is $8\times 10^{-8}$. In Fig. \ref{fig:reach}, we show the region of parameter space $(m_{A'}, \epsilon)$ that can be 
 explored by the experiment with 7 day-equivalent experiment at PAL PLSII. This translates into approximately 6 million beam pulses on target.
We expect approximately 0.5 events due to backgrounds. The figure shows the region where we expect to see 10 signal events or more
in 7 days of data collection. Obviously, for a more realistic performance of an experiment, one has to carefully optimize the optics and choose a suitable photo-detector considering its quantum efficiency and other characteristics. Since the Cherenkov photons would form a circular image, if multi-anode photodetectors are used, additional rejection of backgrounds may be 
possible

\section{Conclusion}
We propose a novel method of detecting the dark photon that decays into a pair of electron
and positron, based on Cherenkov radiation of highly relativistic charged particles in gas.
With this technique, dark photon search becomes possible in the challenging phase space with 
low power electron accelerators due to extremely low backgrounds.
This has been demonstrated for a conceptual experiment using 3 GeV electron LINAC at PAL PLSII in Korea. 
Since no active detector material or geometry is needed, realization of such an experiment
would be simple and low cost. This method could augment the existing methods to enhance the signal-to-background
fraction.

\section{Acknowledgements}
This work has been supported by the Korean National Research Foundation grant 2011-0016554.


\begin{thebibliography}{99}
\bibitem{Pospelov} B. Batell, M. Pospelov, and A. Ritz, Phys. Rev. D, 80 (2009) 095024.
\bibitem{review} M. Raggi and V. Kozhuharov, Riv.Nuovo Cim. 38 (2015) 449.
\bibitem{Bjorken} J. D. Bjorken, R. Essig, P. Schuster, and N. Toro, Phys. Rev. D 80 (2009) 075018.
\bibitem{HPS} HPS Collaboration, Journal of Physics: Conference Series 556 (2014) 012064.
\bibitem{nitrogenscint1} H. Mori et al., Nucl. Instrum. Meth. A 526 (2004) 399.
\bibitem{nitrogenscint2} T. Waldenmaier, J. Blümer, H. Klages, Astropart. Phys. 29 (2008) 205.
\bibitem{PALPLS} http://pal.postech.ac.kr/paleng/
\bibitem{geant4} S. Agostellini et al., Nucl. Instrum. Meth. A 506 (2003) 250.
\end{thebibliography}
\end{document}